\documentclass[a4paper,11pt]{article}
\usepackage{pos}

\title{The Pion Parton Distribution Puzzle in the High x Region}
\author*{Misak Sargsian}
\affiliation{Department of Physics, Florida International University,\\
  11200 SW 8th Street, , Miami, FL, 33193,  USA}

\emailAdd{sargsian@fiu.edu}
\abstract{We developed a new theoretical approach in the description of partonic structure of hadrons based 
on spectral function representation of valence quarks. In this approach massless valence quarks are considered 
in the mean field of residual hadron. This approach initially was applied to the calculation of 
parton distribution function~(PDF) of nucleons that resulted in an observation of pion-cloud structure of nucleon. In this 
report  we present our calculation of  pion PDF in the valence region within the same approach.
In this approach we describe the soft partonic structure of pion through  
the valence $q\bar q$ cluster, embedded in the residual field of the pion.
The pion PDF  is calculated using  phenomenological Light-Front  wave functions for the  $q\bar q$ 
cluster and the residual field.  Our result indicates that the peak position of the  x weighted valence PDF~(xPDF)  
depends on parameters characterizing the virtuality of the $q\bar q$ cluster and the mass of the residual system. 
These parameters are evaluated   by fitting the calculated PDF to the height and the peak position of  empirical pion xPDF at starting $Q_0$. 
They indicate that unlike the nucleon case, very little  residual mass is needed in the pion  to describe existing pion 
PDFs in the valence region. They  also require   the $q\bar q$-cluster to be  highly virtual with 
the interacting quark carrying almost all of the momentum of the $q\bar q$ cluster. 
This picture is consistent with  the dominance of the Feynman mechanism leaving little room for the hard component in the PDF.
The calculation practically describes the  $\sim (1-x)$ behavior observed recently at $x\to 1$ limit. 
Our non-trivial observation is that the  height and the peak position of xPDF define the analytic behavior of valence PDF at $x\to 1$ limit.
}

\FullConference{The 33rd International Workshop on Deep Inelastic Scattering and Related Subjects (DIS2026)\\
4 - 8 May 2026\\
Bologna, Italy\\}

\begin{document}
\maketitle

\section{Introduction}
\vspace{-0.4cm}
As the lightest hadron and a pseudo-Goldstone boson of chiral symmetry breaking, the pion
exhibiting apparent duality: in one hand  mediating   long range part of the nucleon-nucleon~(NN) interaction 
through the near-massless Goldstone field without apparent  internal structure and on the other hand exhibiting 
partonic structure in deep- inelastic scattering~(DIS) processes.  This apparent duality indicates on possible specific distribution 
of valence quarks in the pion that makes for example the quark-interchange in NN  interaction through the  pion field highly unlikely.

The parton distribution function~(PDF) represents an important quantity for exploration of   
QCD dynamics of  the hadron.
%  valence quark distribution in the  pion.  
%It allows  to probe valence and sea quarks as well as gluon 
%distributions each components dominating   at different regions of $x$ in the pion.  
For the valence quarks pion is the best testing ground.
% for understanding the 
%non-perturbative QCD 
%mechanism of valence quark distribution  in the hadrons. 
This fact attracted many theoretical studies of pion ranging from Nambu-Jona-Lasino models, QCD sum rules.
light-front (LF) approaches, lattice calculations and effective LF Hamiltonian with holographic QCD confinement (for the  review of theoretical works  see Ref.\cite{Holt:2010vj,Maerovitz:2025txk}).
Significant progress was made also in the extraction of pion PDFs  exploring 
variety of deep-inelastic reactions including Drell-Yan, $J/\Psi$- production and Sullivan processes 
(see, Refs.\cite{Holt:2010vj,Accardi:2009br,H1:2010hym,Barry:2018ort,Barry:2021osv})
The recent high quality analysis of Drell-Yan and Sullivan processes\cite{Barry:2018ort, Barry:2021osv} produced a {\em puzzling} result  for  PDFs at  large  $x$, observing a behavior of  $(1-x)^\beta$, with $\beta\approx 1-1.2$.  
This contradicts to the generally expected dependence of $(1-x)^2$ which in part follows from the 
expectation that high x parton distributions are generated through the hard gluon exchange between valence quarks\cite{Brodsky:1974vy,Farrar:1975yb,Gunion:1983ay}.

Valence  
%quark distributions  in hadrons 
PDFs are characterized 
by  two important properties, which is used in our calculation: (I) 
The flavor of valence quarks is conserved, thus they  can be considered as effective fermions whose number is conserved.
(II)  Their PDFs weighted by the momentum fraction $x$, $xf_V(x)$
has a  peak associated with the average momentum fraction 
carried by valence quarks.  In general, such a peaking property in  Fermi systems is associated with the ``bulk" property of  bound system which has an inherently non-perturbative character.
In our recent works\cite{Leon:2022isx,Leon:2020cev,Leon:2020bvt} on modeling valence quark  distributions in the nucleon 
the ``bulk" property of hadrons have been described  
through the quark-gluon system residual to the valence quarks in the nucleon. 
This achieved by introducing Light-Front wave function of the
valence quark cluster embedded in the interacting residual field.
 Introducing a residual field in the description of partonic dynamics of the hadron,
which is characterized by its  mass and energy, theoretically corresponds to the spectral function approach in the description of the composite bound system. As such this approach is different from commonly considered theoretical approaches in which valence quarks are the entities that fully describe the partonic dynamics in the intermediate region of $x$,  carrying all the light-front momenta of the nucleon.

Within this model for the case of the nucleon it was found\cite{Leon:2020cev,Leon:2020bvt} that the description of the peaking property of $xf_V(x)$ distribution requires a residual system to have a mass comparable with the pion mass. Thus the spectral function approach  indirectly favored the existence of the structure in  the nucleon resembling a pion cloud.  
Another nontrivial observation within this approach was  that at the 
starting $Q^2_0= m_c^2$ 
%($m_c$ is the mass of the charm quark) 
%when one expects 
%that the partonic  picture to be relevant,  
the peaking position $x_p$
of $xf_V(x)$ distribution satisfies an inequality  $x_p\le {1\over 4}$ (not  $x_p = {1\over 3}$ as naively expected) 
independent to the parameters of the model.
This inequality is  confirmed by the modern phenomenological PDFs of nucleon. 
%Overall, the model described  reasonably well the valence PDFs of nucleons for the region $0.2 < x <0.6$ underestimating  at larger x which indicates   the importance of hard mechanism in the generation of large $x$ component of valence PDFs in the nucleon. 

In the current work we extended the residual field approach for the description of 
pion valence PDFs.   Our goal is to address following main questions by describing the peaking propery of $xf_V(x)$distribution  of pion: 
(I) are there  structure in the pion residual to $q\bar q$ valence quarks (like it was observed for the case of nucleon)  and (II)  how much 
the mean field distribution of the residual system  can contribute to  the high x structure of pion PDFs?.

\section{Theoretical Framework of Residual Field Model}\label{theoretical-framework}
\vspace{-0.4cm}
The residual field model utilizes conservation of valence parton quantum numbers to treat them as effective fermions whose numbers are conserved. It separates hadrons  into a core of valence fermions consisting of massless quarks and a residual structure which contains everything besides the valence quarks, such as sea quarks and gluons. 
Interaction with the external probe in this case proceeds through the 
scattering off the  valence quarks in the cluster. This picture is assumed to be valid for moderate to high $x \gtrsim 0.1-0.2$ 
and will break down  at lower $x$, where the Regge dynamics dominate  in the interaction.

\begin{figure}[htbp]
    \centering
  \includegraphics[width=0.75\linewidth, height=4.cm]{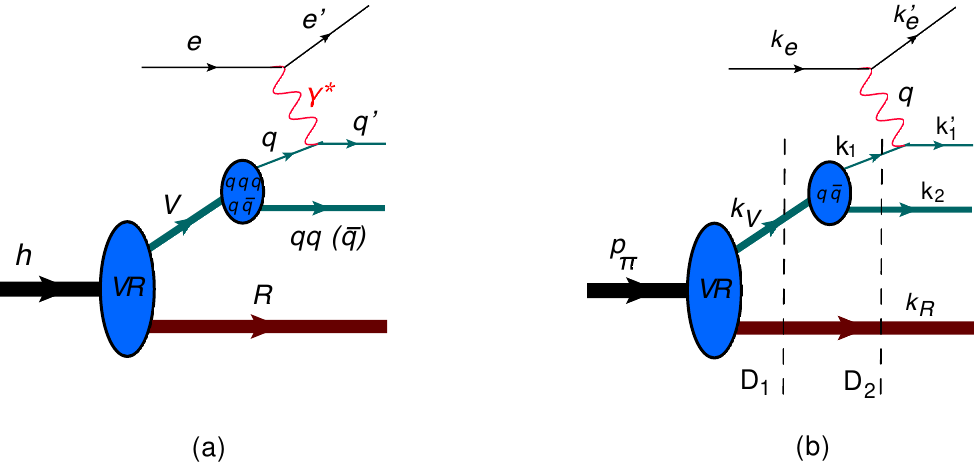}
    \caption{Feynman Diagrams for Deep Inelastic Scattering in Residual-Mean Field model.
    (a) General case for hadrons with $qqq$ or $q\bar q$  valence clusters. (b) For the case of deep inelastic scattering from pion.}
    \label{fig:rfm_diagrams}
    \vspace{-0.4cm}
\end{figure}

According to the residual field model the DIS from a hadron in the leading approximation can be described by the LF-diagram (Fig.(\ref{fig:rfm_diagrams})(a)) in which scattering takes place with a quark inside the valence cluster. The consideration  of the residual system in the scattering amplitude is equivalent  to the spectral function representation of the partonic structure of the hadron in which case the valence quark distribution is indirectly related to the dynamical characteristics of  the residual system.

%Our approach in this work is phenomenological, Furthermore, we define phenomenological wave functions 
%in such a way that they represent the solution of the Weinberg type equation on the light-front for $q\bar q$ - cluster state (V), as well as  for cluster-residual (VR)  effective two body state.
%Such wave functions are inherently non-perturbative and need to be modeled based on 
%the prominent features of the structure of hadrons.

    \vspace{-0.2cm}

\vspace{-0.4cm}
 
\section{The Pion in the Residual Field Model}
\label{pion-in-RMFM}
\vspace{-0.4cm}
To set-up derivational framework for pion PDF we consider probing the pion through 
DIS  ($e \pi \rightarrow e X $). Using the universality of PDFs  the results of calculation can be compared with PDFs extracted for example from Drell-Yan processes in $\pi N\to l^+l^-+X$ reactions. 
%This is a consequence of the definition of PDFs and validity of QCD factorization for these processes. 

We consider  the Drell-Yan-West reference frame where the 4-momenta of the pion, $p_\pi^\mu$ , and virtual photon, $q^\mu$ are defined as follows:
\begin{equation}
p_\pi^\mu = \left(p_\pi^+, \frac{m_\pi^2}{p_\pi^+}, \mathbf{0}^\perp \right), \ \ \ q^\mu = \left(0 , \frac{2 p_\pi\cdot q}{p_\pi^+}, \mathbf{q}^\perp\right),
\label{kins}
\end{equation}
where $q_\perp^2 = Q^2$. 
 In  this reference frame the $F_2$ structure function of the pion is related 
to  the $++$ component of the hadronic tensor as follows:
\vspace{-0.2cm}
\begin{equation}
\vspace{-0.2cm}
F_2(x,Q^2)  =  {m_\pi Q^2\over 2 x (p_\pi^+)^2}W_\pi^{+,+} = \sum\limits_i e_i^2 x f_i(x,Q^2),
\label{F2Wmunu_pdf}
\end{equation}
here $f_i(x)$ are the PDFs of type $i$ partons.  The RHS part of the eqation is valid 
for the definition of PDFs in  leading order.   As follows from above equation   
  PDFs are  related to the   $++$ component of  the hadronic tensor, $W_\pi^{\mu\nu}$.

%\subsection{Hadronic Tensor,  Scattering Amplitude and valence PDF of pion}
For calculation of pion PDF we consider the diagram of 
Fig.\ref{fig:rfm_diagrams}(b) in which it is assumed that the pion transitions to 
the $q\bar q$ cluster  and the residual state, $R$  before 
the  $\gamma^*q(\bar q)$ takes place.

Introducing the  amplitude of $\gamma^*\pi$ scattering, $A^\mu$, (Fig.\ref{fig:rfm_diagrams}(b))  and taking into account the phase space,
  $[dx] \left[d^2 \mathbf{k}_\perp \right]$ , of final state quarks and residual system one obtains for the hadronic tensor: 
\begin{equation}
W^{\mu\nu} (x, Q^2) = \frac{1}{4m_\pi} \sum_{ \{h_i, \tau_i \}} \int  [dx][d^2 \mathbf{k}_\perp]  {2x_1x_B \over Q^2 }\delta(x_1 - x_B),
A^{\mu \dagger} A^\nu.
\label{wmunu3}
\end{equation}
 For calculation of $A^\mu$ (Fig.(\ref{fig:rfm_diagrams})(b))  we used 
LF diagrammatic rules and introduced phenomenological LF wave function for  $q\bar q$ valence cluster, $\psi_{q\bar q}$ and residual system $\psi_{VR}$ resulting in: 
\begin{eqnarray}
 A^+(k_1^\prime,h_1, k_2,-h_1,k_R) = &&   \left[2k_1^+ ie_q
{\psi_{q\bar q}(\beta_1,\beta_2,\mathbf{k}_{1,\perp},\mathbf{k}_{2,\perp})\over \sqrt{2}\beta_1} 
-  2k_1^+ ie_{\bar q}    {\psi_{q\bar q}(\beta_1,\beta_2,\mathbf{k}_{1,\perp},\mathbf{k}_{2,\perp})\over \sqrt{2}\beta_1}\right]
\nonumber \\
&&\times {\psi_{VR}(x_V,\mathbf{k}_{R,\perp},x_R,\mathbf{k}_{V,\perp})\over x_V}.
\label{A+}
\end{eqnarray}

%\subsection{Valence Parton Distribution}
Inserting Eq.(\ref{A+}) into the expression of the hadronic tensor in Eq.(\ref{wmunu3}), summing by helicites of final quarks for   the valence quark PDF one obtains:
\begin{eqnarray}
 f_V(x_B, Q^2) & = &  \int \frac{dx_1 d^2 \mathbf{k}_{1, \perp}}{16\pi^3 x_1} \frac{dx_2 d^2 \mathbf{k}_{2, \perp}}{16\pi^3x_2}  \frac{dx_Rd^2 \mathbf{k}_{R, \perp}}{16\pi^3 x_R}\delta(x_1-x_B)   16 \pi^3 \delta \left(1- x_1-x_2 - x_R\right)
\nonumber \\
& \times &  \delta^{(2)}\left(\sum_{i=1,2,R}\mathbf{k}_{i, \perp}\right)   
\mid  \Psi_{q\bar q}(\beta_1,k_{1,\perp},\beta_2,k_{2,\perp})\mid^2 
\mid \Psi_{VR}(x_R,k_{R,\perp})\mid ^2.
\label{pdf1}
\end{eqnarray}
Furthermore we factorize above integrals by introducing transverse momentum of quark and anti-quark  in the center of mass of the $q\bar q$ system
\begin{eqnarray}
& & 
f_V(x_B, Q^2)   =     
\int_0^{1-x_B} \frac{dx_R }{(16\pi^3)^2 x_B (1 - x_B - x_R )x_R}\nonumber \\
& & \ \ \ \ \ \  \times \left[  
\int^{Q^2} d^2\mathbf{k}_{\perp} |\Psi_{q \bar q}(\beta, {\mathbf{k}}_{\perp})|^2
\right ]_{q\bar q}
\left[\int^{Q^2} d^2 \mathbf{k}_{R, \perp}|\psi_{VR}(x_R, \mathbf{k}_{R, \perp})|^2\right]_{VR}, 
\label{pdf4}
\end{eqnarray}
where $[\cdots]_{q\bar q}$ and $[\cdots]_{V R}$ we can interpret as 
response functions of the $q\bar q$ valence-cluster  and residual systems, respectively.

\section{Modeling Nonperturbative Light-Front Wave Functions}\label{Sec:Model_LFWF}
\vspace{-0.4cm}
To model $q\bar q$ cluster and $VR$ wave functions we  note that both 
are two-body wave functions for which relativistic normalization is defined 
as follows:
\begin{equation}
\int \mid \Psi^{LF}(k_{12})\mid ^2 {dy\over y(1-y)}{d^2k_{12,\perp}\over 16\pi^3} = N,
\label{relnorm}
\end{equation}
where 
$y = {k_1^+\over p_T^+}$,
with $k_1^\mu$ and $p_T^\mu$ being four-momenta of one of the bound particles and 
the target.
Here the relative momentum in the CM frame of two particle system is:
\begin{eqnarray}
&&k_{12}^2 = \frac{\left(s_{12} -(m_1-m_2)^2\right)\left(s_{12}-(m_1+m_2)^2\right)}{4s_{12}}   =  {m_1^2(1-y)+ m_2^2y + k^2_{12,\perp}\over 4 y(1-y)} - {m_1^2+m_2^2\over 2}  \nonumber \\
&&
+ {(m_1^2 - m_2^2)^2y(1-y)\over 4(m_1^2(1-y) + m_2^2 y + k^2_{12\perp})}, \ \mbox{where} \ \ s_{12} = {m_1^2(1-y) + m_2^2y + k^2_{12,\perp}\over y (1-y)}.
\label{relmom}
\end{eqnarray}
 
We model LF wave function similar to \cite{Frankfurt:1981mk,Sargsian:2022rmq} 
assuming factorization of 
the relative momentum  wave function in the CM system.  This allows to relate 
the CM wave function to the non-relativistic 
wave function evaluated at $k_{12}$  defined relativistically in Eq.(\ref{relmom}):
\begin{equation}
\Psi_{12}^{LF}(k_{12}) = \Psi_{12}^{NR}(k_{12}) 
\sqrt{ {E_1E_2\over E_1+E_2}16\pi^3}.
\label{LFWFmodel}
\end{equation}
%Here non-relativistic wave function is normalized as follows:
%\begin{equation}
%$\int \mid \Psi^{NR}(k_{12})\mid ^2 d^3k_{12} = N$.
%\label{NRnorm}
%\end{equation}
With this approach for the LF wave function of the $q\bar q$ cluster one obtains:
 \begin{equation}
\Psi^{LF}_{q\bar q}(\beta,k_{12,\perp}) = \sqrt{8\pi^3 E_k}\Psi^{NR}_{q\bar q}(k_{12}), \ \mbox{where} \ 
 E_k = \sqrt{m^2 + k_{12}^2} \ \ \mbox{and} \ k_{12} = \sqrt{{m^2 +  k^2_{12,\perp}\over 4 \beta(1-\beta)} - m^2}.
\label{wfqqbar}
\end{equation}
%where, $E_k = \sqrt{m^2 + k_{12}^2}$  and 
%$\vec k_{12} = (k_{12,z},\vec k_{12,\perp})$ is the relative momentum 
%in the $q\bar q$ CM system, with the magnitude
%\begin{equation}
%\label{k12}
%\end{equation}
and parameterize the non-relativistic wave function as follows: 
%\begin{equation}
$\Psi^{NR}_{q\bar q}(k) = A_V \exp{\left[-\frac{B_V}{2} k_{12}^2\right]}$.
%\label{qbarqNR}
%\end{equation}

In a similar way we define LF wave function of the residual system as:
\begin{equation}
\Psi_{R}^{LF}(x_R, k_{R,\perp}) = 
A_{R}\sqrt{16\pi^3}\sqrt{ {E_VE_R\over E_V+E_R}}e^{-{B_{R}\over 2} k_{R}^2}, \ \mbox{where} \ E_V = \sqrt{m_V^2 + k_{R}^2}  \ 
\mbox{and}  \ E_R = \sqrt{m_R^2 + k_{R}^2}.
\label{LFVR}
\end{equation}
%where $E_V = \sqrt{m_V^2 + k_{R}^2}$, $E_R = \sqrt{m_R^2 + k_{R}^2}$. 
The $k_R$ is the 3-dimensional momentum defined in the center of mass of the $V-R$ system:

%\subsection{Modeled Soft  Pion PDF}
 Inserting Eqs.(\ref{wfqqbar}) and (\ref{LFVR}) into Eq.(\ref{pdf4}) and performing several integrations analytically one obtains for the pion valence PDF:
\begin{eqnarray}
& & f_V(x_B, Q^2)   =   
%{(\pi) ^{3\over 2}A_V^2 A_{R}^2 \over B_V^{3\over 2} } 
%{(1+h)A_{R}^2}
{N_{q\bar q}A_{R}^2}
 \int_0^{1-x_B} \frac{dx_R }{x_R} {1\over (1-x_R)^2}\ 
 \left[\int^{Q^2} d^2 \mathbf{k}_{R, \perp} 
{E_VE_R\over (E_V+E_R)}e^{-B_R k_{R}^2}  \right]. 
\label{pdf3}
\end{eqnarray}
%The obtained expression indicates that,  the $\beta(1-\beta)$- relative light-cone momentum fraction  
%distribution of valence quarks in the cluster   results in 
%an only  $x_R$ dependent  integrand in  Eq.(\ref{pdf3}), convoluted with the 
%response function of the residual system.
% In the current model this follows from the neglecting the valence quark masses.
%As Eq.(\ref{pdf3}) shows the shape of the PDF depends on the slope factor of the 
%cluster-residual system's momentum distribution, $B_R$, as well as the respective virtual masses of the cluster $m_V$ and the residual %system $m_R$. 

 \vspace{-0.6cm}
 \section{Numerical Results}\label{fitting-results}
 \vspace{-0.4cm}
We use recent results from JAM collaboration\cite{Barry:2018ort}  on 
pion PDF  at starting $Q_0$ to fit Eq.(\ref{pdf3}),  focusing on the description of 
%   Our main focus in the fitting procedure is to
%describe 
the position and the height of the x weighted PDF. 
%in accordance with our 
%model's assumption that these features of valence PDFs are characterized by the soft partonic dynamics. 
%We choose the starting $Q^2= Q_0^2 = 1.69$~GeV$^2$.

The fitting analysis indicates that the position of the peak can be reproduced 
only if the virtual mass of the residual system is close to zero, $m_R\approx 0$.  
This is very different from our analysis of the nucleon PDF's within the same residual field model\cite{Leon:2020cev,Leon:2020nfb} in which we found that  $m_R^N\approx 0.14$, indicating the validity of meson cloud picture  in the nucleon.

\begin{figure}[h]
%     \centering
\hspace{-0.4cm}    \includegraphics[width=0.42\linewidth]{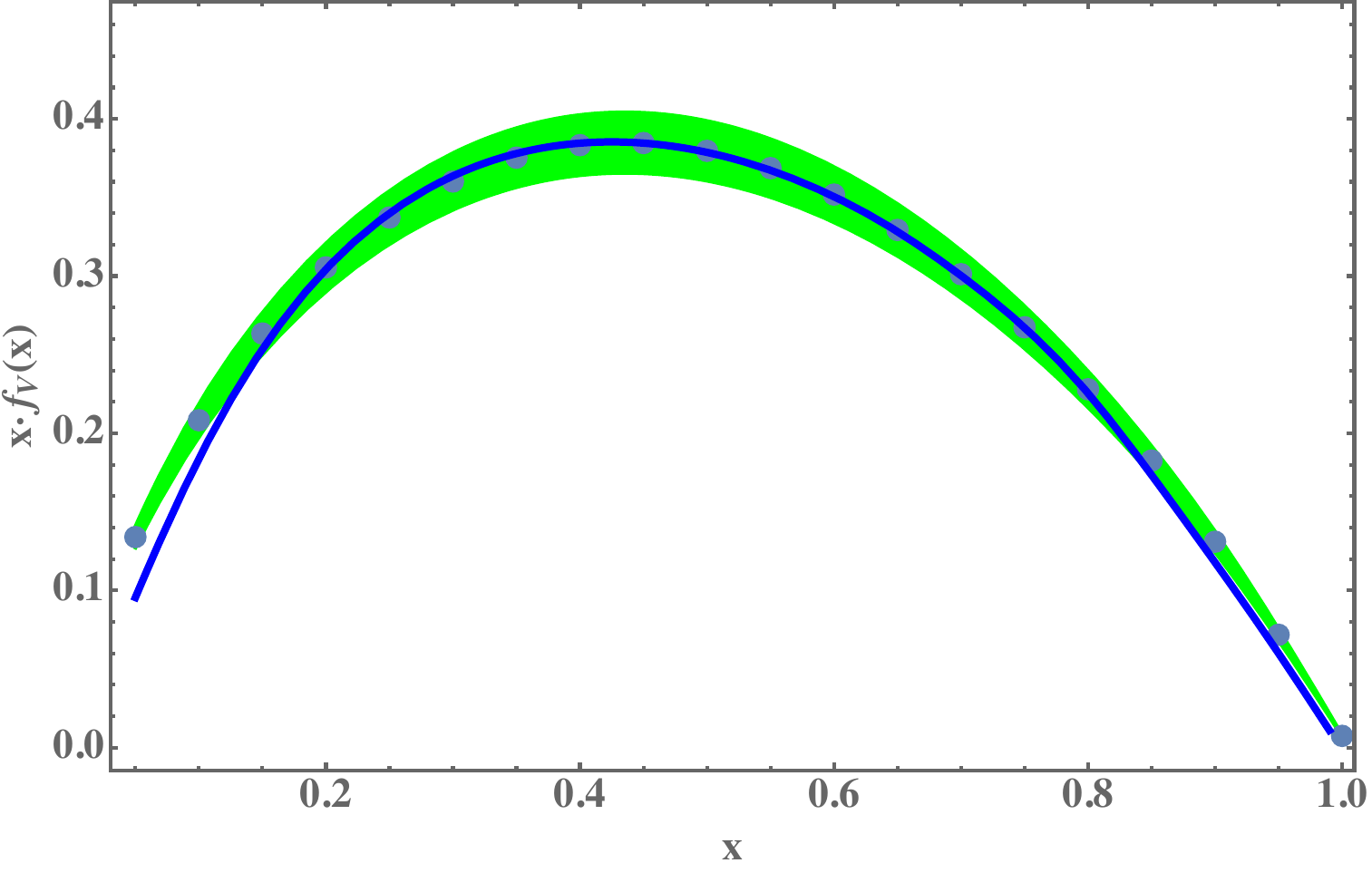}
  \hspace{1cm}
     \includegraphics[width=0.5\linewidth]{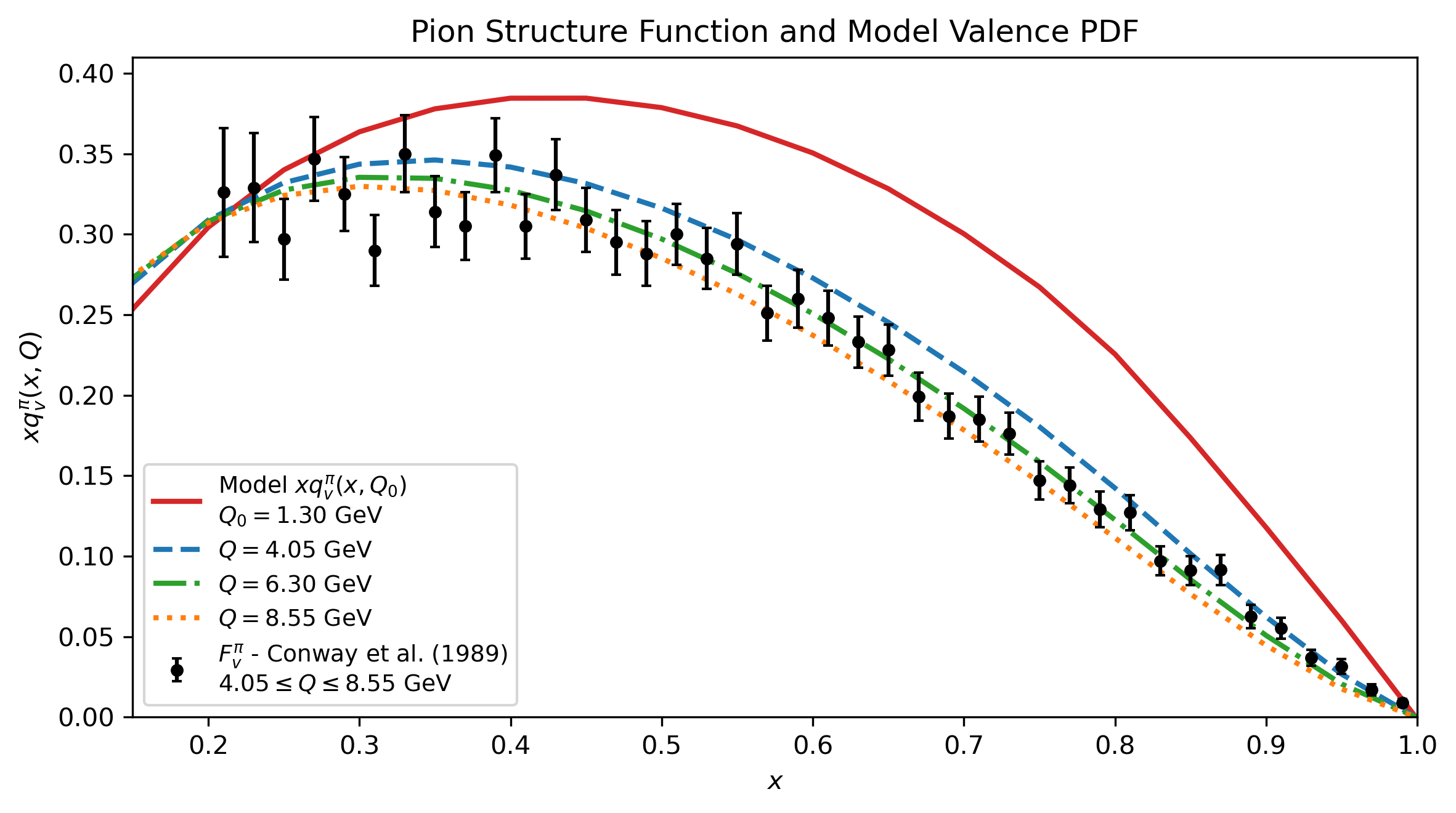}

     \caption{Description of x-weighted pion PDF at $Q_0=1.3$~GeV for $m_R=0$, $m_V=0.95$~GeV and $B_R = 1.2$~GeV$^{-2}$.
     Model pion PDF evolved to different $Q^2$, compared to the experimentally-extracted pion structure functions.}
     \label{fig:two}
 \end{figure}

The best fit of the JAM pion PDF is presented in Fig.\ref{fig:two}(left panel), for the 
parameters of $m_R=0$, $m_V= 0.95$~GeV and $B_V = 1.2$GeV$^{-2}$.  Qualitative agreement of the data requires $m_V\sim Q_0$ and $m_R\sim 0$. As the figure shows, the soft mechanism describes the empirical PDF reasonably well in the range of $0.1< x < 0.85$. 
Even though our goal was not to describe the $x\to 1$ part of the PDF, one observes that our model, predicts a large  contribution to the high x part of the pion PDF   leaving little room for the hard mechanism in the  generation of PDF at large x.

To verify the consistency of our calculation with pion PDFs measured at higher $Q^2$, we compare
our calculation with the PDFs extracted from Fermi-Lab Drell-Yan data\cite{E615:1989bda} on $\pi^-N \to \mu^+\mu^-X$ 
scattering which are measured for dimuon mass range  of $4.05 < Q < 8.55$ GeV, with ($Q \sim m_{\mu\mu}$). 
 For this, we performed QCD evolution of our calculation at starting $Q_0$ to 
the $Q$ values measured at Fermi-Lab. The QCD evolution is performed by using, Evolution Kernel Operators (EKO) library \cite{Candido:2022tld}, which solves the DGLAP evolution equations in Mellin space. 
%The evolution was carried out at leading order, with the one-loop coupling chosen to reproduce $\alpha_s(M_Z) = 0.1180$. 
Fig.\ref{fig:two} shows comparison of evolved calculations for above mentioned 
$Q$ scales with the extracted PDF data. 
%We also present  our model calculation for the initial scale, $Q_0$. 

%  \begin{figure}[h]
%     \centering
 %    \includegraphics[width=0.65\linewidth]{figures/pion_valence_model_input_evolution.png}
%     \caption{Model pion PDF evolved to different $Q^2$, compared to the experimentally-extracted pion structure functions.  }
%     \label{fig:pion_evolution}
 %\end{figure}

As seen in Fig. \ref{fig:two}(right panel),  PDFs evolved from the model calculation at $Q_0$ agrees well with the experimental data. This agreement is consistent with the good  agreement of our model with the pion PDF obtained by JAM collaboration $Q_0$\cite{Barry:2018ort,Barry:2021osv}.
 
 \subsection{Interpretation of the Results}
Since $x$-Bjorken corresponds to the light-front momentum fraction of 
the interacting quark with the relation $x\approx \beta(1-x_R)$, where $\beta$ is 
the LF momentum fraction of the $q\bar q$ cluster carried by the interacting quark, and 
since the broad distribution of $x_R$,
one observes that  for $x\gtrsim 0.5$, relative momentum fraction, $\beta\sim 1$.  
This indicates that  the observed distribution  favors the dominance of the Feynman 
mechanism\cite{Feynman:1973xc} according to which the interacting quark carries the most 
of the light-front momentum fraction  of the cluster, with remaining LF momenta 
distributed among the  wee partons in the residual system.

The observation that $m_R\approx 0$ indicates that  the residual system does not exhibit any structure 
(as it was the case for valence PDFs of nucleons \cite{Leon:2020bvt}). 
On the other hand, the large value of  $m_V\approx 0.95$~GeV makes  the  virtuality of  the 
interacting quark comparable to $Q_0$ of the probe. This together with the  relatively 
large $B_R$ indicates that while interacting valence quark is highly virtual, 
the residual system is spread over the bulk size of the pion.
This picture is consistent with the phenomenology that nucleon-nucleon interactions at large distances ($\ge 1.5fm$) 
proceeds through  the pion exchange in which no quark-interchanges are involved. 

The surprising result is that the considered model leaves little room for 
the hard mechanism of high x PDF.
% generated through the hard gluon exchange between valence $q\bar q$- 
%system. 
It  favors a pion structure dominated by soft mechanism of valence quark interaction.
It is interesting that $f_V(x)\mid_{x\to 1} \sim (1-x)$ consistent with the 
$f_V(x) \sim 1-x$ distribution at the intermediate region in which case $xf_V(x) \sim x(1-x)$ peaks at $x_p \sim 0.5$ in agreement  with the data. %The latter reinforces the observation that high $x$ structure of pion PDF has large soft component that defines also the peaking properties of $xf_V(x)$ distribution.

\section{Conclusion and Outlook}\label{conclusion-and-outlook}
\vspace{-0.4cm}
We have calculated pion valence parton distribution within  residual field model focusing on the intermediate region  of $x$ 
where the $x$ weighted PDF has peaking property. 
Our result  indicates that one can describe the  attributes of pion PDF only if 
the mass of the residual system is close to zero. However, we obtain large virtual mass for 
the $q\bar q$ cluster comparable to the $Q_0$ of the probe, indicating 
% Analysis of the response function 
%of the residual system indicates 
that the dominant mechanism is similar to the Feynman mechanism.
in which interacting parton carries almost all the momentum of the valence quark system, with the remaining light-front momentum distributed to the non-interacting wee partons.

The fit allows to describe the empirical PDF reasonably well in the range of $0.1\le x \le 0.85$
leaving little room for the hard mechanism of generating PDFs at large x.   
  Even though our calculation underestimates the empirical PDF at $x\ge 0.85$, its
substantial contribution in this region  indicates that the inclusion of the hard gluon exchange 
mechanism in generation of parton PDFs at $x\to 1$ will require the consideration of the interference between soft and hard amplitudes that can introduce the exponent, $\beta<2$ in $(1-x)^\beta$ distribution at $x\to 1$. Note that the  $\beta=2$ prediction is based on the assumption of 
negligible contribution from the soft amplitude in the $x\to 1$ region (See e.g. Ref.\cite{Gunion:1983ay}). 

Finally, with the parameters of light-Front wave function of residual system fixed, 
one will be able to make numerical predictions for different observables such as electromagnetic form-factors,  generalized parton distribution functions (GPDs) or transverse momentum dependent parton distribution functions (TMDs) in pion. These  will allow an additional  verification of the model. Additionally, the modeled residual state wave function allows to 
evaluate the fragmentation of pion in the backward kinematic region that can be probed in the EIC collider kinematics.

{\bf Acknowledgment:} This work is supported by United States Department of Energy grant under contract DE-FG02-01ER41172.  

\vspace{-0.2cm}
%\bibliography{pion}

\end{document}